# High-speed free-space quantum key distribution with automatic tracking for short-distance urban links

Alberto Carrasco-Casado [1], María-José García-Martínez [2],
Natalia Denisenko [2], Verónica Fernández [2]

1. Carlos III University, Avda. Universidad 200, Madrid 28911, Spain
2. Spanish National Research Council (CSIC), Calle Serrano 144, Madrid 28006, Spain

Contact name: Verónica Fernández Mármol (veronica.fernandez@iec.csie.es).

**ABSTRACT:**

High-speed free-space quantum key distribution located in urban environment offers an interesting alternative to public key encryption – whose security strength is yet to be mathematically proven. To achieve this, three main objectives need to be accomplished: both the emitter and receiver have to be capable of transmitting and receiving at high speed – with the selection of the source's wavelength and detectors being of especial importance – the error rate needs to be kept at a minimum, especially that due to solar background radiation; and finally, a fast automatic tracking system, capable of compensating for atmospheric turbulence effects, is needed. Regarding to sky background and atmospheric turbulence, two different tracking techniques involving the beam wander compensation in the emitter or the receiver are presented and one of them is selected for our system, based on the link propagation distance and the atmospheric turbulence regime.

## 1.- Introduction

Quantum key distribution (QKD) [1] – or more generally, quantum cryptography – has become a new paradigm in data protection. The laws of quantum mechanics offer a very strong alternative for ensuring data communications. QKD allows two parties to share a cryptographic key using Heisenberg's uncertainty principle as their principal ally. The novelty of this scheme is that the 'quantumness' of this technique allows the legitimate users to detect the presence of an eavesdropper in the channel and thus certify the security of the transmission.

Free-space QKD was primarily aimed towards satellite communications and the main efforts have concentrated in achieving long distances to proof its feasibility [2]. However, short distance (inter-city range) free-space QKD links in urban areas are also of considerable importance, since they can partly alleviate the existing 'connectivity bottleneck'. Furthermore, free-space optics (FSO) in general, has a considerable advantage over optical fiber, namely their flexibility of installation and portability. Unlike optical fiber, that becomes a sunk cost when the customer leaves, FSO can be moved to different locations as required. QKD applied to short free-space links in urban areas is an interesting alternative to current public-key cryptography, which is threatened by a quantum computer attack. In this context, QKD is aimed principally to financial, government and military institutions located within the same city. However, for QKD to be a realistic alternative it needs to be implemented at high speed and have a suitable automatic tracking system that corrects for fast beam deviations caused by atmospheric turbulence.

In this paper, we will discuss how high-speed QKD can be achieved, followed by the experimental system that has been developed at the Information Security Institute (ISI), and the automatic high-speed tracking that is being designed for compensating fast atmospheric fluctuations.

## 2.- Achieving high speed QKD

The first parameter that must be considered when choosing the wavelength source for a free-space QKD system is how well is transmitted through the atmosphere. Usually, there are two spectral windows commonly used in FSO, which are $\lambda \sim 850$ nm and $\lambda \sim 1550$ nm. The atmospheric transmission is higher for the latter (see Fig. 1), especially in urban areas with higher concentration of aerosols, which scatters preferably shorter wavelengths. However, single-photon detectors are also to be considered and in fact, play a dominant role in the selection of the source's wavelength. InGaAs single-photon detectors have good detection efficiencies at $\lambda \sim 1550$ nm, but the frequency they can be operated at they can achieve is limited. Although their performance has improved considerably in the last few years, Si detection technology still outperforms them. Another choice is superconducting detectors, which have excellent timing performance; however, they still have lower detection efficiencies than Si detectors and need to be operated at cryogenic temperatures. Therefore, Si single-photon detectors and a wavelength of 850 nm were chosen as the most efficient combination to achieve GHz- clocked QKD [3].

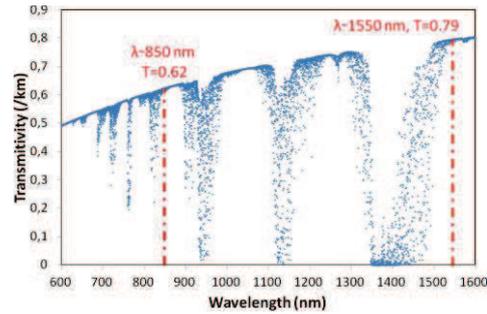

*Fig. 1: Atmospheric spectral transmission in the near-IR for a path of 1 km with urban aerosol and visibility of 5 km. Results simulated using MODTRAN.*

## 3.- The free-space QKD system

The QKD emitter and receiver (figures 2 and 3, respectively, commonly known as Alice and Bob in cryptography) were located at a distance of 300 meters at the CSIC campus in the city center of Madrid. The system implemented the B92 polarization-encoding protocol [4]. The emitter was located at the Institute of Agricultural Sciences (ICA) and the receiver at the Information Security Institute (ISI), both from CSIC.

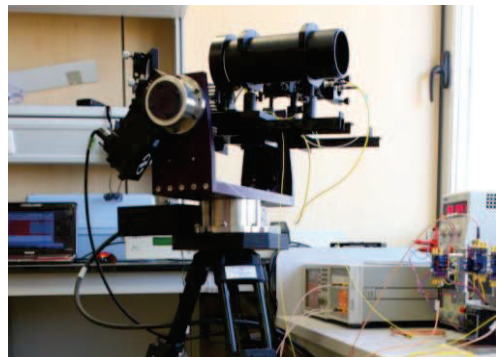

*Fig. 2: emitter of the QKD system.*

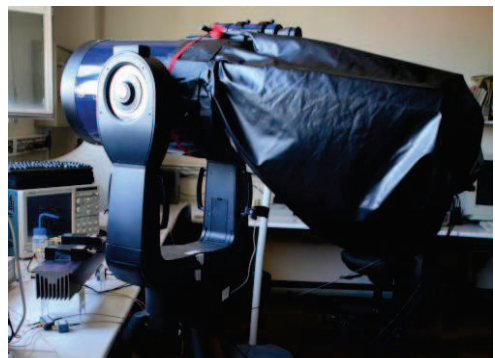

*Fig. 3: Receiver of the QKD system.*

The emitter used a GHz pulse pattern generator to provide a pre-programmed electronic sequence to drive two Gbps vertical-cavity surface-emitting lasers ($VCSEL_0$ and $VCSEL_1$ in fig. 4). They were used to encrypt the binary data of the cryptographic key by using their polarized emission at a relative angle of 45º, and it is usually referred to as the quantum signal. Since this signal cannot be amplified due to the still immaturity of quantum repeaters, losses in the transmission channel and the receiver need to be minimized. Thus, a different signal was used to implement the timing synchronization between emitter and receiver, instead of using a fraction of the quantum signal. Therefore, a third laser ($VCSEL_S$) emitting at $\lambda \sim 1550$ nm was used for timing synchronization between emitter and receiver. This emission was not attenuated to a single photon regime since no key data was encrypted with it. The optical output from $VCSEL_0$ and $VCSEL_1$ was coupled into single mode fibers at a wavelength of $\lambda \sim 850$ nm and connected to fiber-coupled attenuators, which both spatial filters and attenuates the optical beam to secure levels for QKD ($\mu \sim 0.1$ photons per pulse). The output of each fiber is then launched into collimators and coupled by using a 50/50 beamsplitter. A broadband pellicle beamsplitter was used to combine the key data at $\lambda \sim 850$nm with the timing data at $\lambda \sim 1550$nm. The three beams were emitted through a telescope which expanded and collimated the beam to ~40 mm diameter.

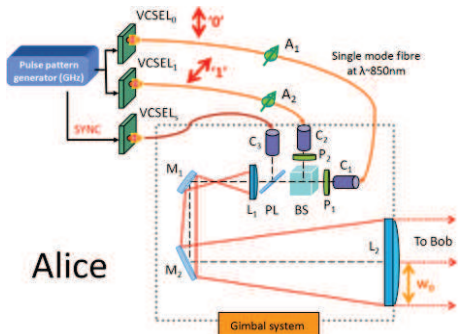

*Fig. 4: QKD emitter. $A_1$ and $A_2$ are attenuators; $P_1$ and $P_2$ are high-extinction ratio polarizers; $C_1$, $C_2$ and $C_3$ are optical collimators, BS is a 50/50 beamsplitter, PL is a pellicle beamsplitter; $L_1$ and $L_2$ are achromatic doublet lenses and $M_1$ and $M_2$ are mirrors.*

The photons reaching the receiver (see fig. 5) are focused by using a Schmidt-Cassegrain telescope and spectrally discriminated by a dichroic beamsplitter, which transmitted the $\lambda \sim 850$ nm 'data' beam, and reflected the $\lambda \sim 1550$ nm synchronization beam. The $\lambda \sim 850$ nm photons are then splitted into two ways indiscriminately and two polarizers are oriented in each channel to correctly select the photons of each state.

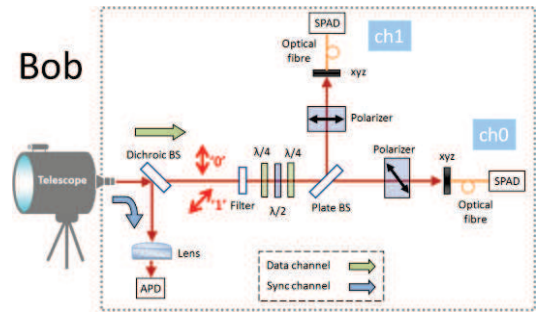

*Fig. 5: QKD receiver. APD is avalanche photodiode; SPAD is single photon avalanche detectors and xyz are traslation stages.*

## 4.- Influence of sky background and turbulence

To assess the security of a key transmission, Alice and Bob analyze the error rate of a small subset of the photon sequence received by Bob. The error rate is commonly referred to as Quantum Bit Error Rate (QBER) and is defined as the number of wrong bits, i.e., bit for which the values of Alice and Bob do not agree over the total number of bits received by Bob. If the QBER is higher than a certain threshold, which depends on the protocol and implementation – for our system is 8 % – the transmission cannot be considered secure. Many factors influence the QBER, such as the noise of the single-photon detectors, polarization imperfections of the quantum states, and especially the solar background radiation, when operated during the day, and atmospheric turbulence. As it will be explained, these two factors present effects that imply confronting mitigation solutions.

To reduce the background noise, a combination of spatial, software and narrow spectral filtering needs to be implemented. In our case, a spectral filter of less than 1 nm combined with spatial filtering by 62.5 μm optical fibers was enough to reduce the background radiation to acceptable levels to enable key rate generation. Fig. 6 shows the effect of the solar background on the QBER and Secret Key Rate (SKR) of the system. As the time of the day increased (sunset was at approximately 21.30 h), the background rate decreased and so did the QBER, causing an increase in the SKR, which reached a maximum value of more than 1 Mbps.

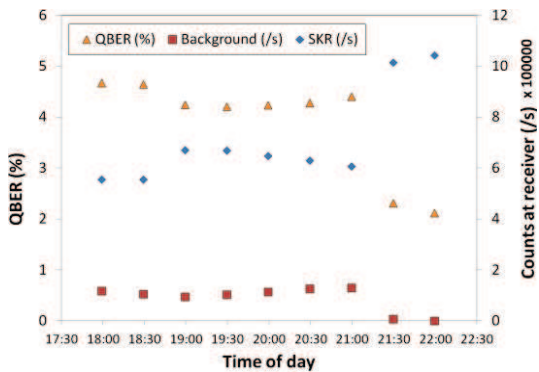

*Fig. 6: Quantum Bit Error Rate (QBER), Secret Key Rate (SKR) and background noise measured at the receiver as a function of the time of the day.*

Atmospheric turbulence is a random space-time distribution of the refractive index, due to air masses movements originated by thermal fluctuations, which in turn affects an optical wavefront in a variety of ways. Those with a more noticeable effect in a QKD optical link are beam spreading and beam wandering. The former is caused by turbulent eddies that are small compared to the beam size, its main effect being an increase in the beam divergence. Beam wander, on the other hand, has its origin in turbulent eddies that are larger than the beam size, with the result of random deflections of the laser beam. Both effects can be combined in the *long-term beam radius*, which models the size that the laser spot would show at the receiver as the result of divergence due to diffraction and beam spreading and the displacement of the beam caused by beam wander over a long time period.

In any free-space lasercom system operating during daylight, a key strategy to limit the sky background that reaches the detector is to minimize the field of view of the receiver's detector. This is especially important in high-speed QKD since this background increases the error rate of the transmission and consequently reduces the achievable bit-rate. The spatial filtering to reject background is achieved by reducing the optical fiber diameter connected to the detector. However, a very narrow field of view is very sensitive to angle-of-arrival fluctuations, which is the very effect of beam wander. These deflections make the signal focuses in different places of the focal plane that could fall out of the optical fiber aperture and therefore cause temporal interruptions of the transmission.

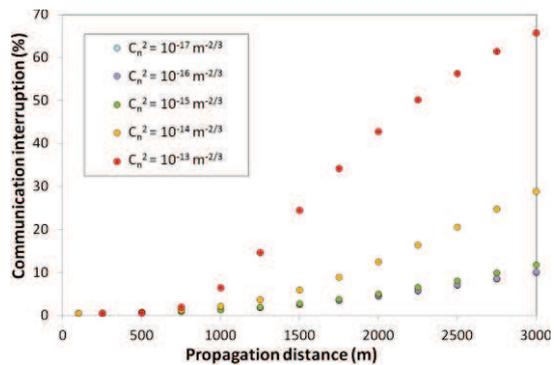

*Fig. 7: Percentage of the transmission where no signal was received due to turbulence.*

Fig. 7 shows the percentage of communication interruption as a function of the distance for various turbulence regimes. The results were simulated for the maximum receiver aperture of the telescope of the QKD system and based on the turbulence model in [5]. The longer the distance, the more noticeable the effect of the turbulence becomes for a QKD system with no tracking. To avoid the significant performance limitations that this approach implies, a tracking system will be implemented to compensate the angle-of-arrival fluctuations.

## 5.- High-speed tracking system

There are a variety of possible setups to achieve the beam wander correction; all of them based on variations of classical laser alignment systems, consisting in a position-sensitive detector (PSD) which data feeds a fast-steering mirror (FSM) closing the loop with a Proportional-Integral-Derivative control (PID). Beam wander is a fast phenomenon, with varying rates that can exceed a hundred Hz, so it is important to design the tracking system to accommodate these fluctuation rates, especially the mechanical parts, namely the fast-steering mirror. In the following, the optical paths from Alice to Bob carrying the quantum and timing information are referred as data and sync channels and the one used to extract the information for beam compensation as the tracking channel.

Beam wander can be modeled as if it was originated from a tip-tilt variation of the laser beam at the transmitter or as an angle-of-arrival fluctuation at the receiver, and these two approaches give rise to two different mitigation techniques. If the long-term beam diameter at the receiver is larger than its aperture size, the compensation should be done at the transmitter. Otherwise, the link would suffer from big losses. Fig. 8 shows a suggested setup to implement this strategy. The goal is to pre-compensate in Alice the beam wander that will affect the quantum channel, based on position measurements of the backwards tracking channel transmitted from Bob at another wavelength. This method is valid for any relation between long-term beam diameter and receiver aperture size, but it implies a more complex setup since an additional laser, possibly with a different wavelength to avoid back reflections coupling into the SPADs. However, this strategy implies a maximum usable distance, since for very long transmission paths, the changes in the atmosphere will be faster than the time involved in the pre-compensation, i.e., the temporal correlation between both atmospheric channels in opposite ways will be lost.

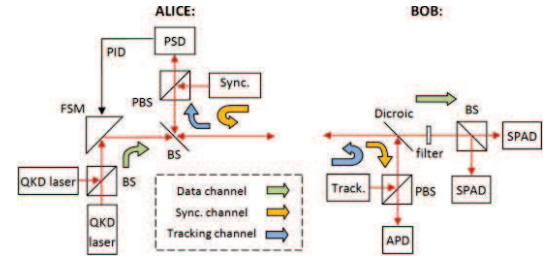

*Fig. 8: Pre-compensation in the emitter.*

In short to medium range paths (not beyond ~3 km in our system), a simpler compensation is possible when the long-term beam diameter is included in the receiver aperture. In this case, since the received spot always enters the telescope, no pre-compensation is needed in Alice, resulting in a more reliable technique (see fig. 9) because there is no limitation in the transmitted distance due to atmosphere correlation in opposite ways. Furthermore, since this scheme is not bidirectional, as was the previous one, the synchronization and the tracking channels can be combined in just one channel, thus excluding one laser from the setup. Now the goal is to correct the angle deviations of the received beam by realigning its position to a pre-established optimum one where the system is perfectly aligned.

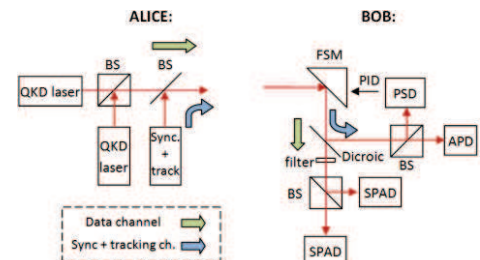

*Fig. 9: Compensation in the receiver.*

Depending on the characteristics of the link, compensation in the receiver could be enough to mitigate beam wander. Otherwise, a pre-compensation in the emitter would have to be used. This will depend on the optical design of Alice and Bob (mainly, the aperture sizes), on the turbulence regime (characterized by the refractive-index structure constant $C_n^2$), and on the propagation distance. A useful way to determine the compensation strategy is to study the receiver aperture to the received long-term diameter ratio. For each turbulence regime, a maximum distance can be determined as a boundary between the two methods (see fig. 10). For this calculation, classic second-order turbulence fluctuation statistics [5] have been used, along with real data from the system described in section 3. It can be observed that at $10^{-15}$ m$^{-2/3}$, which is often referred to average turbulence regime, compensation in the receiver should be used for distances less than 2.45 km. For longer distances, pre-compensation in the emitter should be chosen. For very a strong turbulence regime, ($C_n^2 = 10^{-14}$ m$^{-2/3}$), compensation in Bob would be limited to a distance up to 1.65 km, being the $10^{-13}$ m$^{-2/3}$ value an extremely strong regime, very rarely observed.

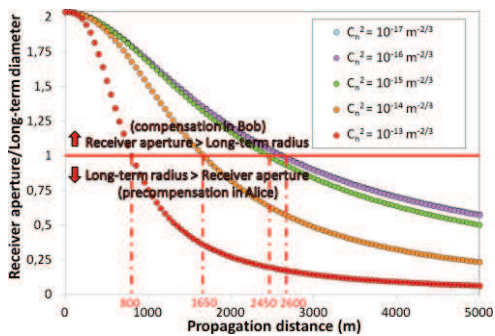

*Fig. 10: Receiver aperture to long-term diameter ratio as a function of distance for several turbulence regimes at a λ ~ 850 nm for a 8cm receiver aperture.*

## 6.- Conclusions

In this paper, the implementation of a GHz-clocked QKD system for urban applications along with a high-speed tracking system that corrects for atmospheric turbulences has been discussed. To achieve high key exchange rates is essential to control critical parameters like the sky background and turbulence. Without tracking, the presented system is capable of achieving secure key rates as high as 1 Mbps – the highest achieved free-space QKD system of these characteristics. Implementing fast automatic tracking could further increase the performance of the QKD system, which involves the compensation of turbulence effects, especially beam wander. Two different compensation techniques have been discussed and one will be implemented to enhance the performance of the QKD system in terms of key exchange rate.

*Acknowledgements*: We thank the project TEC2012-35673 from the *Ministerio de Economía y Competitividad*.